\documentclass[12pt]{article}
\renewcommand{\baselinestretch}{1.35}
\usepackage{graphicx}
\begin{document}
%
\begin{flushright}
  OU-HET-531 \ \\
\end{flushright}
\vspace{0mm}
\begin{center}
\large{Moments of generalized parton distribution functions \\
and the nucleon spin contents}
\end{center}
\vspace{0mm}
\begin{center}
M.~Wakamatsu\footnote{Email \ : \ wakamatu@phys.sci.osaka-u.ac.jp}
\end{center}
\vspace{-5mm}\begin{center}
Department of Physics, Faculty of Science, \\
Osaka University, \\
Toyonaka, Osaka 560-0004, JAPAN
\end{center}

\vspace{-2mm}
\begin{center}
\small{{\bf Abstract}}
\end{center}
\vspace{0mm}
\begin{center}
\begin{minipage}{15.5cm}
\renewcommand{\baselinestretch}{1.0}
\ \ It is shown that, based only on two empirically known facts
besides two reasonable theoretical postulates, we are
inevitably led to a conclusion that the quark orbital angular momentum
carries nearly half of the total nucleon spin.
We also perform a model analysis to find that the quark
spin fraction $\Delta \Sigma$ is extremely sensitive to the pion mass,
which may resolve the discrepancy between the observation and the
prediction of the recent lattice QCD simulation carried out
in the heavy pion region.
\end{minipage}
\end{center}


\vspace{6mm}
 The so-called ``nucleon spin puzzle'' raised more than 15 years ago
is still an unsolved fundamental puzzle in hadron
physics \cite{EMC88},\cite{EMC89}.
If intrinsic quark spin carries little of the total nucleon spin,
what carries the rest of the nucleon spin ? It is the question to be 
answered. Admitting that the QCD is a correct theory of strong interaction,
the answer must naturally be sought for in the following three ;
the quark orbital angular momentum (OAM), the gluon polarization,
and the gluon orbital angular momentum.

Roughly speaking, there exist two contrasting or opposing standpoints to
try to answer the above question. The chiral soliton picture of the nucleon emphasizes the importance of the quark orbital angular
momentum \cite{BEK88},\cite{WY91}.
On the other hand, the possible importance of the gluon polarization
was stressed by several authors in relation with the axial anomaly
of QCD \cite{AR88},\cite{CCM88},\cite{ET88}.
Later, the role of QCD anomaly was understood more clearly within
the framework of the perturbative QCD, especially in view of the
factorization-scheme dependence of parton distribution
functions \cite{BQ90}\nocite{Manohar90}--\cite{BT93}.
Nonetheless, the serious problem is that no one can give any reliable
theoretical prediction for the actual magnitude of $\Delta g$.

An important remark here is that it is meaningless to talk about the
nucleon spin contents without reference to the energy scale of
observation. In fact, it is a widely known fact that the gluon
polarization grows rapidly as $Q^2$ increases, even if it is small at
low energy \cite{JTH96}.
In contrast, the gluon orbital angular momentum decreases
rapidly to partially compensate the increase of $\Delta g$.
(Strictly speaking, these statements are gauge dependent,
since it is known that there is no gauge invariant decomposition of
gluon angular momentum into spin and orbital angular
momentum \cite{JM90}.)
Consequently, when we talk about the nucleon spin contents naively,
we should implicitly understand that we are thinking  of it at low
energy scale of nonperturbative QCD. At this low energy, the CQSM
predicts that \cite{WY91},\cite{WT99},\cite{WW00}
\begin{equation}
 \Delta \Sigma \simeq 0.35, \ \ \ 2 L_q \simeq 0.65,
\end{equation}
which means that the quark OAM dominates over the contribution of quark 
intrinsic spin.

We repeat the question,``Which carries the rest of the nucleon spin,
$L_q$ or $\Delta g$?''
Naturally, only experiments can answer it.
A direct measurement of $\Delta g$ via photon-gluon fusion processes is
one of the most promising direction of study.
For instance, the Compass group
recently extracted the value of $\Delta g / g$ through the analysis of the 
asymmetry of high $p_T$ hadron pairs. Their first result for $\Delta g / g$ has turned out to be  fairly small \cite{Schill05},
\begin{equation}
 \Delta g / g \sim 0.06 \pm 0.31 (stat.) \pm 0.06 (syst.) ,
\end{equation}
although it is premature to draw any decisive conclusion only from this 
analysis.
On the other hand, the key quantity for the direct measurement of $J_q$
or $L_q$ is the generalized parton distributions appearing in the
cross sections of deeply virtual Compton scattering and/or deeply
virtual meson productions \cite{Ji97}\nocite{HJL99}--\cite{Ji98}.
What plays the central role here is Ji's
quark angular momentum sum rule.

Here, we start our argument with the familiar definition of the
generalized form factors $A_{20} (t)$ and $B_{20} (t)$ of the nucleon,
which is given as a nonforward matrix element of QCD energy momentum tensor
$T_{q,g}^{\mu \nu}$ : 
\begin{equation}
 \langle P^{\prime} | T_{q,g}^{\mu \nu} | P \rangle 
 = \bar{U} \,(P^{\prime}) \,
 \left[ A_{20}^{q,g} (t) \gamma^{( \mu} P^{\nu )}
 + B_{20}^{q,g} (t) \,\frac{P^{( \mu} i \sigma^{\nu ) \alpha} 
 \Delta_{\alpha}} {2 M} \,\right] \,U (P) + \cdots .
\end{equation}
According to Ji's sum rule, the total angular momentum carried by quark 
and gluon fields in the nucleon is related to the forward 
($t=0$) limit of these generalized form factors as \cite{Ji98}
\begin{eqnarray}
 J^{u + d} &=& \frac{1}{2} [ A_{20}^{u + d} (0) 
 + B_{20}^{u + d} (0) ], \\
 J^g &=& \frac{1}{2} [ A_{20}^g (0) + B_{20}^g (0) ] .
\end{eqnarray}
Remembering the fact that the above generalized form factors
$A_{20}^{u+d} (0)$ and $A_{20}^g (0)$ are related to the second moments
of the unpolarized generalized parton distribution functions of quarks
and gluons, which reduce to the familiar unpolarized distributions
for quarks and gluons in the forward limit, they just represent the
the total momentum fraction of quarks and gluons in the nucleon as
\begin{eqnarray}
 A_{20}^{u + d} (0) &=& \int_0^1 \,x \,[ u (x) + \bar{u} (x) 
 + d (x) + \bar{d} (x) ] \,dx
 \equiv \langle x \rangle^{u + d} , \\
 A_{20}^g (0) &=& \int_0^1 \,x \,g(x) \equiv \langle x \rangle^g .
\end{eqnarray}
On the other hand, the second $B$ parts are sometimes called the
anomalous gravitomagnetic moments (AGM) of the constituents of
the nucleon \cite{Teryaev99}\nocite{Teryaev98}--\cite{BHMS01}. 
From the conservation of total momentum and angular momentum,
it follows that
\begin{eqnarray}
 A_{20}^{u + d} (0) + A_{20}^g (0) = 1, \\ 
 A_{20}^{u + d} (0) + B_{20}^{u + d} (0) + A_{20}^g (0) 
 + B_{20}^g (0) = 1 ,
\end{eqnarray}
which in turn dictates a nontrivial identity :
\begin{equation}
 B_{20}^{u + d} (0) + B_{20}^g (0) = 0 .
\end{equation}
To proceed further, we must distinguish three possibilities below : 
\begin{enumerate}
 \item $B_{20}^{u + d} (0) = - \,B_{20}^g (0) \neq 0$,
 \item $B_{20}^{u + d} (0) = B_{20}^g (0) = 0$,
 \item $B_{20}^u (0) = B_{20}^d (0) = B_{20}^g (0) = 0$ .
\end{enumerate}
The recent lattice QCD simulation by LHPC Collaboration gives a strong 
support to the second possibility that the total quark contribution
to the nucleon AGM vanishes \cite{LHPC03},\cite{LHPC04}.
This happens as a cancellation of the
$u$- and $d$-quark contributions, i.e., $B_{20}^u (0)$ and
$B_{20}^d (0)$, which have sizable magnitudes with opposite signs.                          
Noteworthy here is the fact that both of $B_{20}^u (0)$ and $B_{20}^d (0)$
have fairly strong dependence on the pion mass but their sum is almost 
independent on it. In any case, this lattice analysis seems to deny the 
third possibility indicated in \cite{Teryaev99} on the basis of the
equivalence principle, but strongly supports the second possibility,
which is the basis of the following argument.
In fact, once we accept this postulate, we are led to 
a surprisingly simple result that the total quark angular momentum is just
a half of the total quark angular momentum
fraction \cite{LHPC03},\cite{LHPC04},\cite{Teryaev99},\cite{Teryaev98} : 
\begin{equation}
 J^{u + d} = \frac{1}{2} \,\langle x \rangle^{u + d} .
\end{equation}
Now, we can go further. First, let us recall an empirically well-accepted
understanding that, even at low energy scale like
$Q^2 \simeq (600 \,\mbox{MeV})^2$, the gluon field seems to carry about
$(20 \sim 30) \%$ of the total nucleon momentum.
(The widespread belief that the quark and gluon fields share equal
amounts of nucleon momentum applies to the asymptotic case of
large $Q^2$). For instance, one may consult the well-established GRV
fit of the unpolarized parton
densities \cite{GRV98}. (See also \cite{GRV95}.)
Their next-to-leading order fit of the gluon density is given
at $Q^2 = \mu_{NLO}^2 = 0.40 \,\mbox{GeV}^2 \simeq (630 \,\mbox{MeV})^2$
as
\begin{equation}
 x g (x, \mu_{NLO}^2) = 20.8 \,x^{1.6} (1 - x)^{4.1} .
\end{equation}
This turns out to give
\begin{equation}
 \langle x \rangle^g \ \equiv \ \int_0^1 x g (x, \mu_{NLO}^2) d x
 \ \simeq \ 0.30 .
\end{equation}
Conversely saying, we can say that, at low energy, the quark field 
carries at least $(70 \sim 80) \%$ of nucleon momentum,
which in turn must be equal to the total quark angular momentum
fraction, according to the aforementioned argument, such that
\begin{equation}
 2 J^{u + d} = \langle x \rangle^{u + d} = (0.7 \sim 0.8) . \label{obs1}
\end{equation}
On the other hand, through the analysis of polarized deep-inelastic 
scatterings, we already know that the intrinsic quark polarization 
$\Delta \Sigma$ is about $(20 \sim 35) \%$ (see, for instance,
the recent review \cite{Bass04}) : 
\begin{equation}
 \Delta \Sigma \simeq (0.2 \sim 0.35) . \label{obs2}
\end{equation}
Putting these two observations (\ref{obs1}) and (\ref{obs2}) together,
we find that the quark orbital angular momentum fraction is
nearly $50 \%$,
\begin{equation}
 2 L^{u + d} = 2 J^{u + d} - \Delta \Sigma \ \simeq \ 0.5.
\end{equation}
That is, once admitting that the isosinglet combination of the quark 
contribution to the nucleon AGM vanishes, we are inevitably led to a 
surprising conclusion that the quark OAM carries nearly half of the
nucleon spin, only with use of the empirically known information.

One might wonder why our conclusion is entirely different from that 
obtained by the LHPC Collaboration \cite{LHPC04},\cite{LHPC04},
who claims that the quark OAM is
negligibly small, in spite that our argument above is based on a result
of the LHPC group, i.e., $B_{20}^{u+d} (0) = 0$.
The rest of the present report is devoted to clarifying this point.
The reason can easily be traced back to the fact 
that, instead of using the empirical value of $\Delta \Sigma$,
they used their theoretical predictions for it,
\begin{equation}
 \Delta \Sigma (\mbox{LHPC}) \simeq 0.682 ,
\end{equation}
which is fairly large and clearly contradicts the observation. Why does 
their analysis give very large $\Delta \Sigma$, then?
This is probably because their simulation was performed with quite large
pion mass around $m_{\pi} \simeq (700 \sim 900) \,\mbox{MeV}$, which is
far from our realistic world close to the chiral limit.
As we shall discuss below, the strong sensitivity of $\Delta \Sigma$
on the pion mass seems to be a likely solution to the above-mentioned
discrepancy.

\vspace{5mm}
\begin{figure}[htb] \centering
\begin{center}
 \includegraphics[width=9.0cm,height=7.5cm]{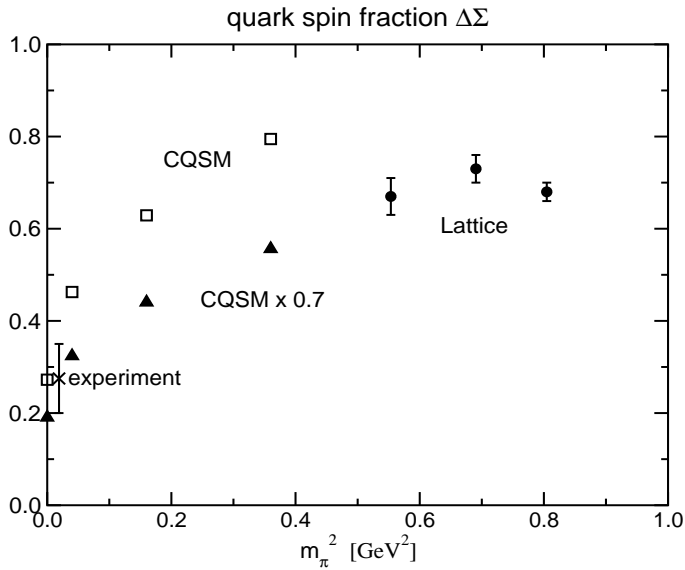}
\end{center}
\vspace*{-0.5cm}
\renewcommand{\baselinestretch}{1.20}
\caption{The quark spin fraction $\Delta \Sigma$ of the nucleon
in dependence of the pion mass $m_\pi^2 \,{[\mbox{GeV}]}^2$.
The filled circles with error bars represent the predictions
for $\Delta \Sigma$ by the LHPC group, corresponding respectively to 
$m_{\pi} = 744, 831$, and $897 \,\mbox{MeV}$, while the cross with error
bar stands for the empirical value corresponding to the physical
pion mass $m_{\pi} = 138 \,\mbox{MeV}$.
The predictions of the CQSM are shown by the open squares for four
values of $m_{\pi}$, i.e., $m_\pi = 0, 200, 400$, and $600 \,\mbox{MeV}$.
Also shown by the filled triangles are the predictions of the
CQSM scaled by the factor 0.7.}
\label{fig:qspin}
\end{figure}%

Now, we shall show it on the basis of the CQSM \cite{DPP88},\cite{WY91}.
Within the framework of the CQSM, we first solve the mean-field
equation of motion self-consistently for several values of $m_{\pi}$.
The model is defined with a physical cutoff.
Here we use the Pauli-Villars regularization scheme with 
double substraction terms \cite{KWW99}. The relevant regularization
parameters are all fixed uniquely from reasonable physical requirements.
How to introduce finite pion mass into the whole scheme is explained
in \cite{KWW99}. Here, we tried to find a self-consistent soliton 
profile with the fixed value of the dynamical quark mass
$M = 400 \,\mbox{MeV}$. This is repeated for several values of pion mass,
i.e., $m_{\pi} = 0, 200, 400$, and $600 \,\mbox{MeV}$.
In this analysis, no stable solution was
found for $m_{\pi} > 620 \,\mbox{MeV}$.
We then evaluate $\Delta \Sigma$ for each soliton solution with different
value of $m_{\pi}$. The results are shown in Fig.1 together with the
predictions of the LHPC Collaboration \cite{LHPC03},\cite{LHPC04}.
The filled circles with error bars represent the predictions
for $\Delta \Sigma$ by the LHPC group, corresponding respectively to 
$m_{\pi} = 744, 831$, and $897 \,\mbox{MeV}$, while the cross with error
bar stands for the empirical value corresponding to the physical
pion mass $m_{\pi} = 138 \,\mbox{MeV}$.
The predictions of the CQSM are shown by the open squares for four
values of $m_{\pi}$, i.e., $m_\pi = 0, 200, 400$, and $600 \,\mbox{MeV}$.
One clearly sees that $\Delta \Sigma$ is very sensitive to the value of
$m_{\pi}$, especially when approaching the chiral limit
$m_{\pi} \rightarrow 0$. 
Inspired by the indication of the GRSV fit, which dictates that the quark
fields carries only $70 \%$ of the total nucleon momentum and also the
total spin, one may tentatively renormalize the predictions of the CQSM by
multiplying a factor of 0.7.
The results are shown by the filled triangles.
It is interesting to see that these points appears to be smoothly
connected to the lattice predictions given in the large $m_{\pi}$ domain.
We hope that the lattice simulation in the near future will be extended to
the region of smaller $m_{\pi}$ and that it will confirm the strong 
$m_{\pi}$ dependence of $\Delta \Sigma$ predicted by the CQSM, although
the lattice QCD would still need a help of some other theoretical
technique like the chiral perturbation theory to explore the region
of $m_{\pi}$ very close to the chiral limit \cite{YLT04}.

Summarizing our arguments, we have shown that, with use of the two
empirical knowledge alone, aside from the two reasonable theoretical
postulates, we are inevitably led to a drastic conclusion that the
{\it quark orbital angular momentum carries nearly half of the total
nucleon spin}. The two theoretical postulates here are
\begin{itemize}
 \item Ji's angular momentum sum rule \ : \ 
 $J^{u + d} = \frac{1}{2} \,[ \langle x \rangle^{u + d} + 
 B_{20}^{u + d} (0) ]$ ,
 \item absence of the {\it net quark contribution} to the
 {\it anomalous gravitomagnetic moment} of the nucleon \ : \
 $B_{20}^{u + d} (0) = 0$ .
\end{itemize}
On the other hand, the two empirically known facts we have used are
\begin{itemize}
 \item the fraction of the quark momentum and angular momentum of the
 nucleon at low  energy scale, $Q^2 \simeq (600 \,\mbox{MeV})^2$ \ : \ 
 $\langle x \rangle^{u + d} = 2 J^{u + d} \simeq (0.7 \sim 0.8)$ ,
 \item the quark spin fraction from polarized DIS analyses \ : \ 
 $\Delta \Sigma \simeq (0.2 \sim 0.35)$ .
\end{itemize}
Although there remains some room concerning how to define the angular
momentum of the constituents of the nucleon \cite{BJ98},
it is reasonable to stick to Ji's definition,
which leads to the above sum rule. Otherwise, we would 
lose a only clue to experimentally access the quark angular momentum in
the nucleon. Thus, only one factor, which might potentially alter our
conclusion, is the second postulate, i.e., $B_{20}^{u + d} (0) = 0$.
Although it is strongly supported by the lattice simulation by the
LHPC Collaboration, an independent check is highly desirable.
Also desirable is an analytical proof of it within the framework
of (nonperturbative) QCD.

We have also shown that the above-mentioned conclusion, {\it obtained
independently of any models}, is qualitatively consistent with the 
predictions of the CQSM. The CQSM predicts very strong dependence of
the quark spin fraction $\Delta \Sigma$ on the pion mass : it reproduces
small $\Delta \Sigma$ in the domain close to the chiral limit, it also
smoothly matches the predictions of the LHPC Collaborations obtained in
the heavy pion region. It is hoped that this behavior of $\Delta \Sigma$
will be confirmed by the lattice simulation in the near future.

\vspace{8mm}
\begin{flushleft}
\begin{Large}
{\bf Acknowledgement}
\end{Large}
\end{flushleft}
\vspace{3mm}
This work is supported in part by a Grant-in-Aid for Scientific
Research for Ministry of Education, Culture, Sports, Science
and Technology, Japan (No.~C-16540253)


\begin{thebibliography}{10}

\bibitem{EMC88}
EMC Collaboration~: J.~Aschman~et al.
\newblock {\em Phys. Lett.}, B206:364, 1988.

\bibitem{EMC89}
EMC Collaboration~: J.~Aschman~et al.
\newblock {\em Nucl. Phys.}, B328:1, 1989.

\bibitem{BEK88}
S.J. Brodsky, J.~Ellis, and M.~Karliner.
\newblock {\em Phys. Lett.}, B206:309, 1988.

\bibitem{WY91}
M.~Wakamatsu and H.~Yoshiki.
\newblock {\em Nucl. Phys.}, A524:561, 1991.

\bibitem{AR88}
G.~Altarelli and C.G. Ross.
\newblock {\em Phys. Lett.}, B212:391, 1988.

\bibitem{CCM88}
D.~Carlitz, J.C. Collins, and H.A. M\"{u}ller.
\newblock {\em Phys. Lett.}, B214:229, 1988.

\bibitem{ET88}
A.V. Efremov and O.V. Teryaev.
\newblock {\em JINR Report}, E2-88:287, 1988.

\bibitem{BQ90}
G.T. Bodwin and J.~Qiu.
\newblock {\em Phys. Rev.}, D41:2755, 1990.

\bibitem{Manohar90}
A.V. Manohar.
\newblock The g(1) problem : much ado about nothing.
\newblock In {\em University Park Workshop}, page~90. 
AIP Conf. Proc. 223, 1991.

\bibitem{BT93}
S.D. Bass and A.W. Thomas.
\newblock {\em J. Phys.}, G19:925, 1993.

\bibitem{JTH96}
X.~Ji, J.~Tang, and P.~Hoodbhoy.
\newblock {\em Phys. Rev. Lett.}, 76:740, 1996.

\bibitem{JM90}
R.L. Jaffe and A.~Manohar.
\newblock {\em Nucl. Phys.}, B536:303, 1990.

\bibitem{WT99}
M.~Wakamatsu and T.~Kubota.
\newblock {\em Phys. Rev.}, D60:034020, 1999.

\bibitem{WW00}
M.~Wakamatsu and T.~Watabe.
\newblock {\em Phys. Rev.}, D62:054009, 2000.

\bibitem{Schill05}
C.~Schill.
\newblock {\em hep-ex/0501056}, 2005.

\bibitem{Ji97}
X.~Ji.
\newblock {\em Phys. Rev. Lett.}, 78:610, 1997.

\bibitem{HJL99}
P.~Hoodbhoy, X.~Ji, and W.~Lu.
\newblock {\em Phys. Rev.}, D59:014013, 1999.

\bibitem{Ji98}
X.~Ji.
\newblock {\em J. Phys.}, G24:1181, 1998.

\bibitem{Teryaev99}
O.V. Teryaev.
\newblock {\em hep-ph/9904376}, 1999.

\bibitem{Teryaev98}
O.V. Teryaev.
\newblock {\em hep-ph/9803403}, 1998.

\bibitem{BHMS01}
S.J. Brodsky, D.-S. Hwang, B.-Q. Ma, and I.~Schmidt.
\newblock {\em Nucl. Phys.}, B593:311, 2001.

\bibitem{LHPC03}
Ph. H\"{a}gler, J.W. Negele, D.B. Renner, W.~Schroers, T.~Lippert, and
K.~Schilling.
\newblock {\em Phys. Rev.}, D68:034505, 2003.

\bibitem{LHPC04}
J.W. Negele, R.C. Brower, P.~Dreher, R.~Edwards, G.~Fleming,
Ph. H\"{a}gler, U.M. Heller, Th. Lippert, A.V. Polchinsky, D.B. Renner,
D.~Richards, K.~Schilling, and W.~Schroers.
\newblock {\em Nucl. Phys.}, B128:170, 2004.

\bibitem{GRV98}
M.~Gl\"{u}ck, E.~Reya, and A.~Vogt.
\newblock {\em Eur. Phys. J.}, C5:461, 1998.

\bibitem{GRV95}
M.~Gl\"{u}ck, E.~Reya, and W.~Vogelsang.
\newblock {\em Phys. Lett.}, B359:201, 1995.

\bibitem{Bass04}
S.~Bass.
\newblock {\em hep-ph/0411005}, 2004.

\bibitem{DPP88}
D.I. Diakonov, V.Yu. Petrov, and P.V. Pobylitsa.
\newblock {\em Nucl. Phys.}, B306:809, 1988.

\bibitem{KWW99}
T.~Kubota, M.~Wakamatsu, and T.~Watabe.
\newblock {\em Phys. Rev.}, D60:014016, 1999.

\bibitem{YLT04}
R.D. Young, D.B. Leinweber, and A.W. Thomas.
\newblock {\em Nucl. Phys. Proc. Suppl.}, 128:227, 2004.

\bibitem{BJ98}
S.V. Bashinsky and R.L.~Jaffe.
\newblock {\em Nucl. Phys.}, B536:303, 1998.

\end{thebibliography}
\end{document}